\documentclass[12pt, prd, showpacs]{revtex4}
\usepackage{amsmath}
\usepackage{amssymb}

\input{tcilatex}

\begin{document}

\title{Static black holes in equilibrium with matter: nonlinear equation of
state}
\author{Oleg B. Zaslavskii}
\affiliation{Astronomical Institute of Kharkov V.N. Karazin National University, 35
Sumskaya Street., Kharkov, 61022, Ukraine}
\email{ozaslav@kharkov.ua}

\begin{abstract}
We consider a spherically symmetric black hole in equilibrium with
surrounding classical matter that is characterized by a nonlinear dependence
of the radial pressure $p_{r}$ on the density $\rho $. We examine under
which requirements such an equilibrium is possible. It is shown that if the
radial and transverse pressures are equal (Pascal perfect fluid), equation
of state should be approximately linear near the horizon. The corresponding
restriction on $\frac{dp_{r}}{d\rho }$ is a direct generalization of the
result, previously found for an exactly linear equation of state. In the
anisotropic case there is no restriction on equation of state but the
horizon should be simple (nondegenerate).
\end{abstract}

\maketitle

%Use showkeys class option if keyword display desired

% It is always \today, today, but any date may be explicitly specified

%\newpage

\section{Introduction}

In real physical circumstances, any black hole is surrounded by matter. If
some special conditions imposed on the equation of state are fulfilled, an
equilibrium between a black hole and matter can occur, so that there is no
collapse, the metric being static or stationary. As a result, some sort of
hair appears around a black hole. If matter is described by the linear
equation of state, only quite definite discrete set of such a hair
enumerated by two integers is possible \cite{c}. Meanwhile, in general
nonlinear equations of state look more realistic. Therefore, of interest is
the question: which types of matter are compatible with the static black
hole horizon in the nonlinear case? Below, we suggest an answer for the
simplest but physically relevant case of spherically symmetric
configurations, thus extending the results of \cite{c}.

Consider a spherically symmetric metric which can be written in the form%
\begin{equation}
ds^{2}=-Adt^{2}+\frac{du^{2}}{A}+r^{2}(d\theta ^{2}+\sin \theta ^{2}d\phi
^{2}).
\end{equation}

Here $u$ is called a quasiglobal coordinate, and the location of the event
horizon corresponds to the outmost zero of $A$. It is convenient to use the
combination of 00 and 11 Einstein equations, the conservation law and 11
equation. They read, respectively,

\begin{equation}
2A\frac{r^{\prime \prime }}{r}=-8\pi (p_{r}+\rho ),  \label{01}
\end{equation}%
\begin{equation}
p_{r}^{\prime }+\frac{2r^{\prime }}{r}(p_{r}-p_{\perp })+\frac{A^{\prime }}{%
2A}(p_{r}+\rho )=0,  \label{cl}
\end{equation}

\begin{equation}
\frac{1}{r^{2}}(-1+A^{\prime }rr^{\prime }+Ar^{\prime 2})=8\pi p_{r}
\label{11}
\end{equation}%
where the prime denotes derivation with respect to $u$. \ Without the loss
of generality, we can assume that on the horizon $u=0$. We are interested in
the conditions (if any) of which equation of state should obey for the
configuration to be regular in the vicinity of the horizon. The latter
means, by definition, that the metric coefficients and the energy density
should be analytical functions of $u$ in the vicinity of the horizon. In
particular, near the horizon of the n-th order $A\sim u^{n}$ where $n>0$ is
an integer.

\section{Equation of state: general asymptotic form}

We assume the general form of the relationship between the density and
radial pressure%
\begin{equation}
p_{r}=f(\rho )\text{.}
\end{equation}

First of all, we are going to elucidate which general asymptotics of $f(\rho
)$ is admissible near the horizon. On the horizon itself we must have \cite%
{fn} (p. 207), \cite{vis}%
\begin{equation}
\rho _{h}+p_{hr}=0
\end{equation}%
where the subscript "h" means that the corresponding quantity is calculated
on the horizon. Let now for $\rho \rightarrow \rho _{h}$%
\begin{equation}
f(\rho )\approx -\rho _{h}+w_{h}(\rho -\rho _{h})^{\alpha }  \label{as}
\end{equation}%
where $w_{h}$ and $\alpha >0$ are constants.

In what follows, the properties of the system depend crucially on whether or
not $p_{r}-p_{\perp }=0$. In the linear case $f=w\rho $ considered in \cite%
{c} it was unimportant. Indeed, because of the assumption $\left\vert
p_{\perp }\right\vert /\rho <\infty $ \cite{c} (which is physically
reasonable and is more weak than the energy dominant condition) the simple
estimate shows that the second term in (\ref{cl}) has the same order as $%
\rho $ and can be neglected near the horizon where $\rho \rightarrow 0$ (see 
\cite{c} for details). Meanwhile, now $\rho _{h}\neq 0$ in general, so one
should be careful. Therefore, we consider separately the isotropic case 
\begin{equation}
p_{r}=p  \label{sym}
\end{equation}%
(which corresponds to the perfect fluid) and the anisotropic one $p_{r}\neq p
$.

\section{Isotropic case}

Now, the second term in (\ref{cl}) cancels. The equation can be integrated
to give 
\begin{equation}
A=A_{0}\exp (-2z),z=\int \frac{d\rho }{f+\rho }\frac{df}{d\rho }.
\label{int}
\end{equation}

If $\alpha >1$, one can obtain from (\ref{int}) or (\ref{cl}) and (\ref{as})
that near the horizon $z\sim (\rho -\rho _{h})^{\alpha -1}+const$, $A\approx
const\neq 0$ contrary to the fact that on the horizon we must have $A(\rho
_{h})=0$. Thus, matter with the equation of state under discussion having $%
\alpha >1$ cannot support a horizon.

If $\alpha <1$, we obtain%
\begin{equation}
A\sim (\rho -\rho _{h})^{-2\alpha }\text{.}
\end{equation}

Instead of the horizon, we have a singularity, so this choice of $\alpha $
also should be rejected.

Thus, only the case $\alpha =1$ is admissible: near the horizon any equation
of state should behave as a linear one.

Let now $\alpha =1$. Then, we assume also that near the horizon 
\begin{equation}
\rho -\rho _{h}=B_{k}u^{k}+o(u^{k})
\end{equation}%
where $k>0$ is an integer, and $B_{k}$ is a constant. Then, collecting in (%
\ref{cl}) the main contributions of each term which have the order $k-1$ and
equating the corresponding coefficient to zero, we obtain that%
\begin{equation}
w_{h}=\left( \frac{df}{d\rho }\right) _{\rho =\rho _{h}}=-\frac{n}{n+2k}<0%
\text{, }\left\vert w_{h}\right\vert <1  \label{w}
\end{equation}%
that generalizes the previous observation made in \cite{c} for the
particular case $f=w\rho $ with a constant $w$.

The same holds true if, instead of the exact equality (\ref{sym}), we have
near the horizon%
\begin{equation}
\left\vert p_{r}-p_{\perp }\right\vert \leq C\left\vert \rho -\rho
_{h}\right\vert  \label{sym1}
\end{equation}%
where $C$ is some constant.

\section{Anisotropic case}

Now we assume that the conditions (\ref{sym}), (\ref{sym1}) are violated.

As we want to derive some general model-independent conclusions, we suppose
that, generically, $r_{h}^{\prime }\neq 0$. Then, in eq. (\ref{cl}) the term
of the zero order appears due to the difference $p_{r}-p_{\perp }$. To
satisfy eq. (\ref{cl}) and compensate this term, we must have $k=1$, with%
\begin{equation}
\frac{B_{1}}{2}[w_{h}(n+2)+n)]+\frac{2r_{h}^{\prime }}{r_{h}}(p_{r}-p_{\perp
})_{\mid u=u_{h}}=0
\end{equation}%
One can determine the coefficient $B_{1}$ from this equation, so there is no
restriction on the form of equation of state. Moreover, the quantity $w_{h}$
may have any sign, so this case includes matter with $\frac{\partial p_{r}}{%
\partial \rho }>0$ that can be of interest bearing in mind the stability
issue. Further, as $k=1$ and it follows from (\ref{01}) that $k\geq n$
(assuming also that $r_{h}^{\prime \prime }\neq 0$), we have also $n=1$ (a
simple horizon).

\section{Vacuum fluid and near-horizon expansion: comparison}

Near the horizon, taking the expansion $\rho =\rho _{h}+\tilde{\rho}$ and $%
p_{r}=-\rho _{h}+\tilde{p}_{r}$ with small $\tilde{\rho}$ and $\tilde{p}_{r}$%
, one can represent approximately the total stress-energy tensor as%
\begin{equation}
T_{\mu }^{\nu }=\tilde{T}_{\mu }^{\nu }+T_{(vac)\mu }^{\nu }  \label{vac}
\end{equation}%
where $\tilde{T}_{\mu }^{\nu }$ is linear with respect to $\tilde{\rho}$ and 
$p_{_{(vac)}r}=-\rho _{vac}\equiv -\rho _{h}$. Thus, by definition, $%
T_{(vac)\mu }^{\nu }$ has the form of so-called dark (vacuum) fluid.
Therefore, it is instructive to compare the results for 1) a generic
nonlinear equation of state and case 2) when from the very beginning there
is a mixture of two noninteracting terms - the matter with the linear
equation of state and the vacuum fluid \cite{c} \ In the isotropic case $%
p_{r}=p_{\perp }$ (perfect fluid) the results are the same, with (\ref{w})
slightly generalizing the corresponding relation for the linear equation of
state \cite{c}. However, if $p_{r}\neq p_{\perp }$ they are qualitatively
different. Namely, in case 2 eq. (\ref{w}) still holds whereas in a case 1
there is no restriction on the equation of state at all. Instead, there is a
restriction on the type of the horizon (only $n=1$ is admissible) in case 1
whereas there is no such a restriction in case 2 \cite{c}. This difference
between the two cases can be attributed to the fact that in case 1 one
cannot split the total stress-energy tensor to two non-interacting parts, so
that the zero--order term contribution the second term in (\ref{vac}) enters
the equation for $\tilde{p}_{r}$ and $\tilde{\rho}$ in (\ref{cl}).

It is also worth noting that in \cite{c} (where $\rho _{h}=p_{hr}=0$) eq. (%
\ref{11}) was used to distinct the possible types of the horizon when the
vacuum field is present or absent. Now, it is irrelevant since, in general, $%
p_{hr}\neq 0$ without any additional vacuum fluid and any type of the
horizon - simple $(n=1$) or degenerate ($n\geq 2$) is possible. For the case 
$n\geq 2$, only the first term in the left hand side of (\ref{11}) survives,
so $p_{hr}=-\frac{1}{8\pi r_{h}^{2}}$.

\section{Conclusion}

Thus, the results of \cite{c} are extended to the arbitrary nonlinear
equations of state. It turned out that the properties of equation of state
necessary for the possibility of matter to be in equilibrium with the
horizon of a static spherically symmetric black hole depend crucially on
whether or not the matter represents the perfect fluid. In the first case,
any admissible equation of state should behave as effectively linear near
the horizon, higher or fractional powers in the expansion like (\ref{as})
are excluded. Further, there is a restriction on the first derivative $\frac{%
dp_{r}}{d\rho }$ on the horizon that generalizes the corresponding result
for the linear case \cite{c} directly. However, if the radial and
transversal pressures do not coincide, the existence of the horizon does not
impose any restrictions on the form of equation of state. Instead, it
restricts strongly the type of the horizon which is shown to be simple only.
Although any stress-energy tensor near the horizon is approximately equal to
the vacuum contribution plus small corrections, the results are
qualitatively different from \cite{c} due to the effective interaction
between both components.

One can rephrase these results by saying that the Pascal perfect fluid
admits only discrete set of hair but does not restrict the order of the
horizon, whereas generic anisotropic matter allows any hair but only simple
horizons.

All restrictions discussed in the present work are model ndependent. In
addition, any concrete model can, obviously, add some special restrictions.

\section{Acknowledgement}

I thank Roman Konoplya for a stimulating question on the seminar that
entailed writing this work. I am grateful to the Institute for Theoretical
Astrophysics of T\"{u}bingen University where this work started, for
hospitality and warm atmosphere.

I also thank Kirill Bronnikov for reading the manuscript and useful comments.

\end{document}